\documentclass{article}
 
\usepackage{amsmath,amssymb}
\usepackage[mathscr]{eucal}
\usepackage{graphicx, color}

\newcommand{\rf}[1]{(\ref{#1})}
\newcommand{\ba}{\begin{array}}
\newcommand{\ea}{\end{array}}

\newcommand{\bracket}[2]{\bra{#1}\,#2\rangle}
\newcommand{\bra}[1]{\langle\,#1\,|}
\newcommand{\ket}[1]{|\,#1\,\rangle}

\newcommand{\p}{\partial}
\newcommand{\ud}{\mathrm{d}}
\newcommand{\pathD}{\!\mathscr{D}}

\newcommand{\pp}{{\bf p}}

\title{QED amplitudes: recurrence relations to all orders}
\date{}
\author{Anton Ilderton \\ \\ Department of Mathematical Sciences \\ University of Durham\\ South Road \\ Durham DH1 3LE, United Kingdom \\  \texttt{a.b.ilderton@dur.ac.uk}}

\begin{document}
\maketitle
\abstract{\noindent We describe the origins of recurrence relations between field theory amplitudes in terms of the construction of Feynman diagrams. In application we derive recurrence relations for the amplitudes of QED which hold to all loop orders and for all combinations of external particles. These results may also be derived from the Schwinger-Dyson equations.}    
\section{Introduction}

There is currently a wealth of interest in the perturbation series of Yang-Mills theory expressed in terms of maximal helicity violating, or MHV, amplitudes. These amplitudes may be used as a set of vertices with which to generate an alternative perturbation expansion of the theory. 

However, until very recently there was a tree level proof that these results hold generally \cite{proof}, with numerous examples at low loop order \cite{loops1} - \cite{loops3}. It has now been shown how to write the Yang-Mills action in terms of variables which makes the MHV structure manifest \cite{Paul}. The quadratic part of the action is the inverse of a scalar propagator which connects positive and negative helicity fields while the infinite number of tree level MHV amplitudes appear as an infinite number of vertices in the action. This is an exciting result since it allows perturbative Yang-Mills to be re-developed in terms of Feynman diagrams constructed from MHV amplitudes. This approach describes amplitudes through recurrence relations, a computationally more efficient approach than the usual Feynman diagram expansion. Examples of similar recursion results have been found to hold in QED and gravity \cite{qed} \cite{gravity}, though only at low orders.

There is, then, considerable interest in alternative perturbative structures in quantum field theory. In this paper we will show that such recursion relations may be derived from properties of correlation functions. We will construct recursion relations for QED amplitudes which hold for all external particles and to all orders in the loop expansion.

This paper is organised as follows. We begin with an example, using scalar fields, which illustrates our method. The example is based loosely on arguments from the sewing of string world sheets and suggests that the object to be calculated is the expectation value of the free action. In section \ref{Scalar fields} we evaluate this in terms of functional operators and use power counting arguments to generate recurrence relations between correlation functions to all loop orders. We explain how this leads directly to recurrence relations between scattering amplitudes. In section \ref{QED} we extend our arguments to QED, deriving recurrence relations for the scattering of photons and fermions. Finally in section \ref{Conclusions} we give our conclusions, summarise an alternative derivation from the Schwinger-Dyson equations and discuss the extension to other theories.
\section{An example}
Consider the four particle scattering amplitude in $\phi^3$ theory, given by the LSZ formula
\begin{equation}\label{one}
  \bracket{\pp_1, \pp_2}{\pp_3, \pp_4} = \int\!\ud x_1\ldots \ud x_4\,\,e^{ip_1.x_1}e^{ip_2.x_2}e^{-ip_3.x_3}e^{-ip_4.x_4} \prod_{j=1}^4 i\big(\p^2_{x_j} + m^2\big)\langle\,1\ldots 4\,\rangle
\end{equation}
where the $p_i$ are on shell. Our metric is mostly minus. We denote an integration over all $D+1$ dimensional spacetime by $\ud x$ and we have abbreviated the connected correlation functions to
\begin{equation*}
\langle\,1\ldots n\,\rangle \equiv \bra{0}\,T\big[\hat\phi(x_1)\ldots\hat\phi(x_n)\big]\,\ket{0}_\text{connected}.
\end{equation*}
In scalar $\phi^3$ theory the tree level contribution to the correlation function above is
\begin{equation}\label{2}
  \langle\,1\ldots 4\,\rangle_0 =  -\lambda^2\sum\limits_\sigma\int\!\ud z\ud z'\,\, G_{x_1 z'}G_{x_2 z'}G_{zz'}G_{z x_3}G_{z x_4}
\end{equation}
where the subscript zero indicates tree level, $G_{xy}$ is the propagator between points $x^\mu$ and $y^\mu$ and the sum over $\sigma$ is over inequivalent permutations of the external points, in this case giving three possible diagrams each with a factor of $-\lambda^2$, below.
\begin{equation*}
  \includegraphics[width=0.4\textwidth]{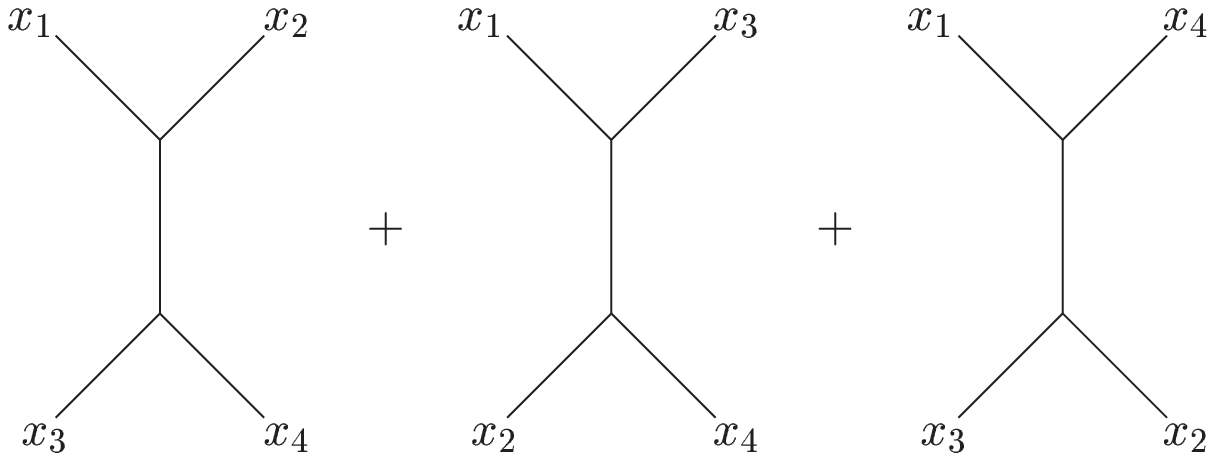}
\end{equation*}
The key observation is that these diagrams may be split into three field diagrams be severing the internal line. This can be performed explicitly using the particle equivalent of Carlip's sewing method for string worldsheets \cite{Carlip}. In the string case two worldsheets are sewn together at a boundary by integrating over shared (co-ordinate and ghost) boundary data but with an insertion of the string Hamiltonian acting on one of the boundaries. This is necessary to correctly cover the moduli space of the worldsheet being formed.

The same consideration must be applied to particle theories, albeit at a basic level. We will be interested in cutting and joining propagators. A careful derivation of the propagator which maintains reparametrisation invariance, see \cite{Polyakov}, reveals a modular parameter which is the intrinsic length of the worldline. This is usually not manifest since the propagator is commonly represented by its Fourier transform. The propagator here is $G_{xy}$, the inverse of $i\delta(x-y)(\p^2_x+m^2)$ (the Gaussian operator in the partition function with an appropriate $i\epsilon$ prescription). For this object Carlip's sewing becomes the almost trivial
\begin{equation}\label{carlip-sew}
  G_{xy} = \int\!\ud z\,\, G_{xz}i(\p^2_z+m^2)G_{zy}.
\end{equation}
Returning to our example we apply \rf{carlip-sew} twice to the internal line, separating the four field diagrams into two three field diagrams. The correlation function can then be written
\begin{equation}
  \langle\,1\ldots 4\,\rangle_0 = \sum\limits_\sigma \int\!\ud a\ud b\,\, \bigg[i(\p^2_a+m^2) \langle\,1,2,a\rangle_0\bigg]G_{ab}\bigg[i(\p^2_b+m^2)\langle\,3,4,b\rangle_0\bigg]
\end{equation}
where the sum over $\sigma$ takes in the symmetry of $G_{ab}$ and therefore the equivalence of the two terms in large brackets. This may be checked using the explicit form of the amplitude in (\ref{2}). The insertions of the Hamiltonian are of course the operators appearing in the LSZ reduction formula. If we insert our result into \rf{one} and write $G_{ab}$ as its Fourier transform,
\begin{equation*}
  G_{ab} = i\int\!\frac{\ud^{D+1}k}{(2\pi)^{D+1}}\,\,\frac{e^{-ik.(a-b)}}{k^2-m^2}
\end{equation*}
which provides the necessary exponential factors, then the four particle scattering amplitude becomes
\begin{equation}
  \bracket{\pp_1, \pp_2}{\pp_3,\pp_4}_0 =\sum\limits_\sigma\int\!\frac{\ud k}{(2\pi)^{D+1}}\,\, \bracket{\pp_1,\pp_2}{k}_0\,\frac{i}{k^2-m^2}\,\bracket{k}{\pp_3, \pp_4}_0
\end{equation}
where the sum is now over inequivalent permutations of the external momenta (including $k$). Since each S matrix element is proportional to a momentum conserving delta function we may perform the integral over $k^\mu$, leaving
\begin{equation}
  \mathcal{A}(p_1,p_2\,|\,p_3,p_4)_0 = \sum\limits_\sigma\mathcal{A}(p_1,p_2\,|\,k)_0\frac{i}{k^2-m^2}\,\mathcal{A}(k\,|\,p_3,p_4)_0
\end{equation}
where $k$ conserves momentum, $k=p_1+p_2 = p_3+p_4$. We have expressed the four particle scattering amplitude in terms of three particle scattering amplitudes connected by a propagator (a structure familiar from, but not the same as, the MHV rules). In the next section we will derive a general form of this result which holds to all loop orders.
\section{Scalar fields}\label{Scalar fields}
Before discussing QED we will give the general result of applying Carlip's method to a scalar field theory with interaction $\lambda\phi^N/N!$ We would like to show that all connected $n$--point correlation functions are related to $n+2$--point correlation functions via the use of \rf{carlip-sew}. Let the partition function for the theory be $Z[J]=e^{W[J]}$ where $W$ is the generator of connected correlation functions. We are interested in proving something like
\begin{equation}\begin{split}
  W[J] &\sim \int\!\ud x\ud y\,\,(\p^2_x+m^2)\frac{\delta W[J]}{\delta J(x)} G_{xy} (\p^2_y+m^2)\frac{\delta W[J]}{\delta J(y)} \\
  &\sim\int\!\ud x\,\,\frac{\delta W[J]}{\delta J(x)}(\p^2_x+m^2)\frac{\delta W[J]}{\delta J(x)}
\end{split}\end{equation}
where the two additional fields on the right hand side become derivatives with respect to the source $J$. The second line in the above follows from an integration by parts and points us at the object we will investigate, namely the expectation value of the free action $S_0$ in the presence of a source, 
\begin{equation}\label{begin}
  \frac{i}{\hbar}S_0\big[-i\hbar\frac{\delta}{\delta J}\big]Z[J] = \int\pathD\phi\,\,\frac{i}{\hbar}S_0[\phi]\exp\bigg(\frac{i}{\hbar}S[\phi] + \frac{i}{\hbar}\int\!J\phi\bigg).
\end{equation}
Introducing a parameter $\zeta$ we can write \rf{begin} as
\begin{equation}\label{ahha}
  \frac{i}{\hbar}S_0\big[-i\hbar\frac{\delta}{\delta J}\big]Z[J] = \frac{\p}{\p\zeta}\bigg|_{\zeta=0}  \int\pathD\phi\,\,\exp\bigg(\frac{i(1+\zeta)}{\hbar}S_0[\phi] + \frac{i}{\hbar}S_N[\phi] + \frac{i}{\hbar}\int\!J\phi\bigg)
\end{equation}
The free scalar field action is 
\begin{equation*}
  S_0[\phi] = \frac{1}{2}\int\!\ud x\,\,\p\phi.\p\phi - m^2\phi
\end{equation*}
so the left hand side of \rf{ahha} is
\begin{equation}
Z[J]\frac{\hbar}{2}\int\!\ud x\,\, i(\partial^2_x + m^2)\frac{\delta^2 W}{\delta J(x)^2} + \frac{\delta W}{\delta J(x)}i(\partial^2_x + m^2)\frac{\delta W}{\delta J(x)}.
\end{equation}
To evaluate the right hand side of  \rf{ahha}, observe that $W$ is the sum of connected diagrams constructed from $V$ vertices, $n$ external lines and $I$ internal lines, with $l$ loops. We expand $W[J]$ as  
\begin{equation}
  W[J] = \frac{1}{\hbar}\sum\limits_{n=0}^\infty \int\!\ud x_1\ldots\ud  x_n\,\,\frac{i^n}{n!}J(x_1)\ldots J(x_n)\langle\,1\ldots n\,\rangle.
\end{equation}
Each of the correlation functions has a loop expansion in powers of $\hbar$,
\begin{equation}
  \langle\,1\ldots n \,\rangle = \sum\limits_{l=0}^\infty \hbar^l \langle\,1\ldots n \,\rangle_l
\end{equation}
with $l=0$ being the tree level pieces etc.  The only change to the generating functional $W$ which results from the inclusion of the $\zeta$ term in the exponential is that each propagator is accompanied by a factor $(1+\zeta)^{-1}$. Therefore each Feynman diagram picks up a factor of $(1+\zeta)^{-n - I}$. Using the Euler formula and the loop counting formula,
\begin{equation}\label{counting}
  n+ 2I = NV, \qquad I-V = l-1,
\end{equation}
we deduce that each set of connected Feynman diagrams which make the correlation functions in the expansion of $W$ now carries a factor 
\begin{equation}\label{factor}
  (1+\zeta)^{-n-I(n,l)}\quad\text{where}\quad I(n,l):= (n+N(l-1))/(N-2).
\end{equation}
Returning to \rf{ahha} the derivative with respect to $\zeta$ brings down this new sum of connected diagrams from the exponential. Setting $\zeta$ to zero leaves $W$ with $-n-I(n,l)$ multiplying the $n$--field $l$--loop correlation function, all multiplied by the usual partition function. Equating both sides of \rf{ahha} we find,
\begin{equation}\begin{split}
\frac{\hbar}{2}\int\!\ud x\,\, &i(\partial^2_x + m^2)\frac{\delta^2 W}{\delta J(x)^2} + \frac{\delta W}{\delta J(x)}i(\partial^2_x + m^2)\frac{\delta W}{\delta J(x)} \\
&=  -\frac{1}{\hbar} \sum\limits_{n=0}^\infty \int\!\ud x_1\ldots\ud  x_n\,\, \frac{i^n}{n!}J(x_1)\ldots J(x_n)\sum\limits_{l=0}^\infty \hbar^l (n+I(n,l))\langle\,1\ldots n\,\rangle_l.
\end{split}\end{equation}
Equating the coefficients of $J^n\hbar^{l-1}$ gives
\begin{equation}\label{nearly}\begin{split}
  (n+I(n,l))\langle\,1\ldots n\,\rangle_l &= \frac{1}{2}\int\!\ud x\,\,i(\partial^2_x+m^2)\langle\,1\ldots n, x,x\rangle_{l-1} \\
  &+ \frac{1}{2}\text{Sym}\sum\limits_{p=0}^n\sum\limits_{l'=0}^l\int\!\ud x\, \langle\,1...p,x\,\rangle_{l'}\frac{i(\partial^2_x+m^2)}{p!(n-p)!}\langle\,p+1\ldots n,x\,\rangle_{l-l'}
\end{split}\end{equation}
where the symmetrisation is over all possible distributions of $x_1$ to $x_n$ in the pair of correlation functions. This is our result. It describes recurrence relations, in the number of particles and the loop order, between the correlation functions of scalar field theory which hold to all loop orders. \\
\newline
To be thorough, it can be seen that when the left hand side is $\langle\,1\ldots n\,\rangle_l$ the right hand side contains terms multiplying correlation functions of $n$ fields and $n+1$ fields, both at loop order $l$; this appears to spoil the recurrence properties. The offending terms come from $p=0$ ($p=n$) and $p=1$ ($p=n-1$). When $p=0$ a tadpole multiplies $(\p^2+m^2)\langle\,1\ldots n, x\,\rangle$, but it is only the tree level tadpole which multiplies $(\p^2+m^2)\langle\,1\ldots n, x\,\rangle_l$, and this term vanishes since by definition there is no tree level tadpole. The right hand side therefore contains correlation functions of $n+1$ fields but of order at most $l-1$.

When $p=1$ there is a tree level two point function multiplying $(\p^2+m^2)\langle\,1\ldots n-1,x\,\rangle_l$, but this becomes a delta function after integration by parts and we can move this contribution to the left hand side. The right hand side then contains correlation functions of $n$ fields but of order at most $l-1$. We therefore have genuine recurrence relations in the number of fields and loop order. Following these details our result becomes
\begin{equation}\label{final}\begin{split}
  I(n,l)\langle\,1\ldots n\,\rangle_l = &\frac{1}{2}\int\!\ud x\,\,i(\partial^2_x+m^2)\langle\,1\ldots n, x,x\rangle_{l-1} \\
  &+ \frac{1}{2}\text{Sym}\sum\limits_{p=2}^{n-2}\sum\limits_{l'=0}^l\int\!\ud x\, \langle\,1...p,x\,\rangle_{l'}\frac{i(\partial^2_x+m^2)}{p!(n-p)!}\langle\,p+1\ldots n,x\,\rangle_{l-l'} \\
  &+ \sum\limits_{1\rightarrow 2\rightarrow\ldots n}\sum\limits_{l'=1}^{l-1} \int\!\ud x\,\,\langle\,1,x\,\rangle_{l'} i(\p^2_x+m^2) \langle\,2\ldots n ,x\,\rangle_{l-l'} \\
  &+ \sum\limits_{l'=1}^{l-1} \int\!\ud x\,\,\langle\,x\,\rangle_{l'} i(\p^2_x+m^2) \langle\,1\ldots n ,x\,\rangle_{l-l'}.
\end{split}\end{equation}
At tree level there are no connected $n$--point functions for $n<N$ and our result is empty in this sector. For $n>N$ we have the simple relation
\begin{equation}\label{tree}
\langle\,1\ldots n\,\rangle_0 =\frac{n-N}{2N-4}\text{Sym}\sum\limits_{p=2}^{n-2}\int\!\ud x\, \langle\,1...p,x\,\rangle_0\,i(\partial^2_x+m^2)\langle\,p+1\ldots n,x\,\rangle_0.
\end{equation}
In all of the expressions above we may insert under each integral a resolution of the identity,
\begin{equation}\label{reso}
  1 = \int\!\ud y\,\, i(\p^2_y+m^2)G_{xy},
\end{equation}
which can be used to extend the recurrence relations to scattering amplitudes. The tree level relations become, for example,
\begin{equation}\label{tree2}
\langle\,1\ldots n\,\rangle_0 =\frac{n-N}{2N-4}\text{Sym}\sum\limits_{p=2}^{n-2}\int\!\ud x\ud y\, \big[i(\p^2_y +m^2)\langle\,1...p,y\,\rangle_0\big]\,\frac{G_{xy}}{p!(n-p!}\,\big[i(\partial^2_x+m^2)\langle\,p+1\ldots n,x\,\rangle_0\big].
\end{equation}
Multiplying by a product of operators $i(\p^2+m^2)$ for each variable and taking the Fourier transform gives, on the left hand side, the expression for the $n$--particle scattering amplitude. For example, setting $N=3$, $n=4$ and $l=0$ we recover the result for the four particle scattering amplitude in $\phi^3$ theory from the previous section.


\section{Quantum electrodynamics}\label{QED}
Our QED Lagrangian in the gauge $\p^\mu A_\mu=0$ is
\begin{equation}
\mathcal{L}=\frac{1}{2}A_\mu\big(\p^2\eta^{\mu\nu}\big)A_\nu + \bar\psi(i\gamma.\p - m)\psi - e\bar\psi \gamma^\mu\psi A_\mu
\end{equation}
which is obtained from the standard action by including the gauge fixing term
\begin{equation*}
  \delta_Q\big[b\p^\mu A_\mu + \frac{i}{2}b\lambda\big]
\end{equation*}
with BRST transformations $\delta_Q A_\mu = \p_\mu c$, $\delta_Q^2 A_\mu=0$, $\delta b =\lambda$, $\delta \lambda=0$. The ghost integrations give an irrelevant factor to the partition function while the integral over $\lambda$ is Gaussian and removes the piece proportional to $\p^\mu A_\mu$ from the term quadratic in the gauge field. As before we will calculate the expectation value of the free action in the presence of sources,
\begin{equation*}
  \int\pathD(A,\psi,\bar\psi)\,\,\bigg[\frac{i}{\hbar}\int\!\ud x\,\,\frac{1}{2}A_\mu\big(\p^2\eta^{\mu\nu}\big)A_\nu + \bar\psi(i\gamma.\p - m)\psi\bigg]e^{iS/\hbar + iS_\text{src}/\hbar}
\end{equation*}
where the source action is
\begin{equation*}
  S_\text{src} = \int\!\ud x\,\, J^\mu A_\mu +\bar\psi_a\eta^a+ \bar\eta^b\psi_b
\end{equation*}
for vector $J$ and Grassmann valued (independent) spinors $\eta$ and $\bar\eta$. Introducing the parameter $\zeta$ the expectation may be written
\begin{equation}\label{forQED}
\frac{i}{\hbar}S_0\bigg[-i\hbar\frac{\delta}{\delta J},-i\hbar\frac{\delta}{\delta \bar\eta},+i\hbar\frac{\delta}{\delta \eta}\bigg] Z[J,\bar\eta,\eta] = \frac{\p}{\p\zeta}\bigg|_{\zeta =0}\int\pathD(A,\psi,\bar\psi)\,\,e^{i(1+\zeta)S_0/\hbar + iS_\text{Int}/\hbar + iS_\text{src}/\hbar}.
\end{equation}
Written in terms of functional derivatives the free action acts on $W$, the generator of connected graphs, as
\begin{equation}\begin{split}\label{big-op}
  &\frac{\hbar}{2}\int\!\ud x\,\,(-i\p^2\eta^{\mu\nu})_x\frac{\delta W}{\delta J^\mu(x)}\frac{\delta W}{\delta J^\nu(x)} + \frac{\hbar}{2}\int\!\ud x\,\,(-i\p^2\eta^{\mu\nu})_x\frac{\delta^2 W}{\delta J^\mu(x)\delta J^\nu(x)} \\ 
  &+ \hbar\int\!\ud x\,\, \frac{\delta W}{\delta \eta^a(x)}(-\gamma.\p_x-im)^{ab}\frac{\delta W}{\delta\bar\eta^b(x)} + \hbar\int\!\ud x\,\,(-\gamma.\p_x-im)^{ab}\frac{\delta^2 W}{\delta\eta^a(x)\delta\bar\eta^b(x)}.
\end{split}\end{equation}
The processes represented by these four terms are illustrated below, respectively.
\begin{equation*}
  \includegraphics[width=0.6\textwidth]{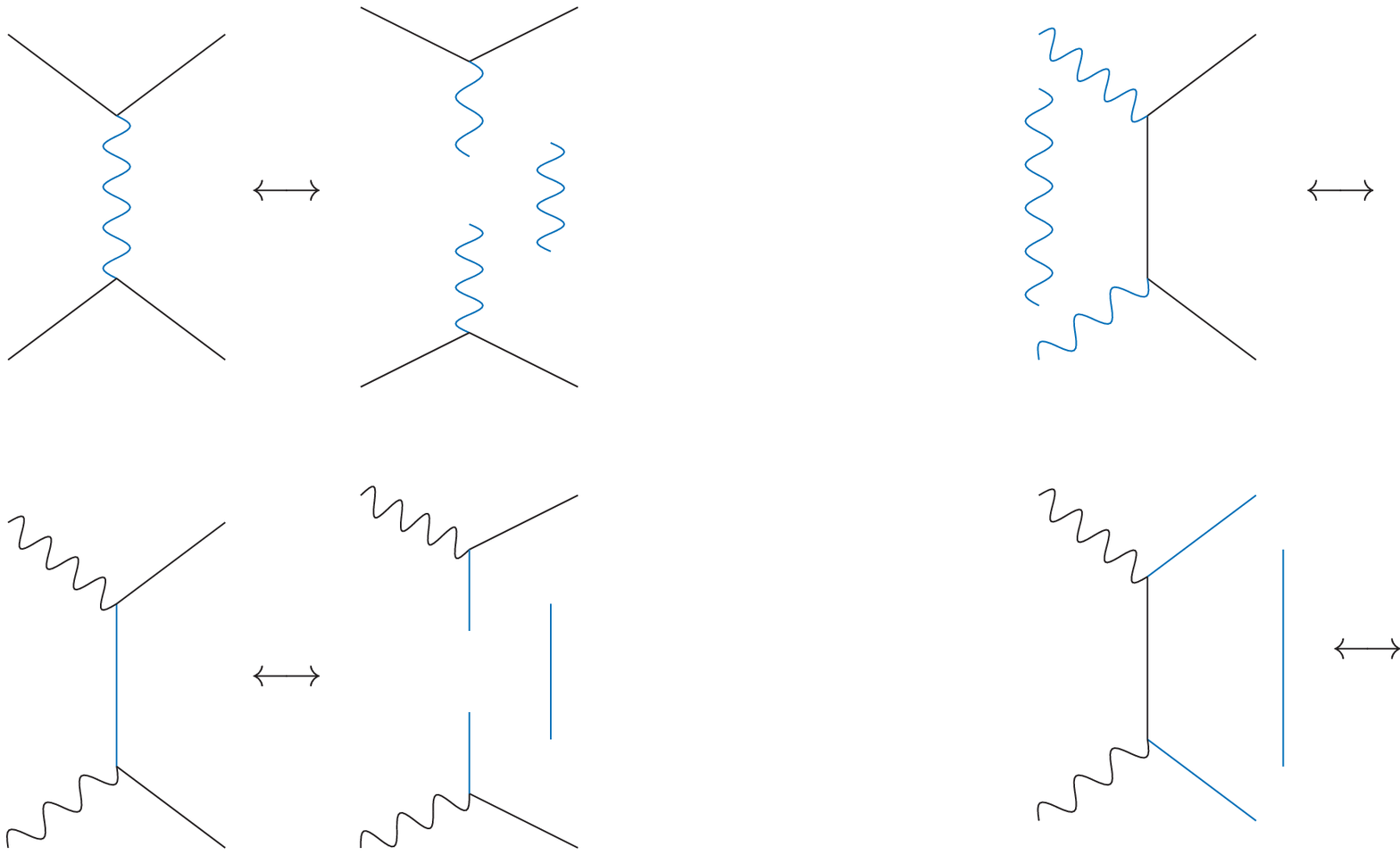}
\end{equation*}
The first order derivatives join two diagrams by joining one external line from each to form a single internal line. Proving results beyond tree level has been one of the challenges of research in MHV and related areas. The terms with second order derivatives, above, deal with the loops in our approach. They act on a single diagram, joining two external lines together to form a loop, or cutting a loop open which decreases an amplitude's loop number by one but adds two additional external legs, thus relating $n$ and $n+2$ particle scattering of one lower loop order. 

As we now have three different fields our equations will become more cluttered. The results will be clearer if we first calculate the recurrence relations between purely bosonic and fermionic amplitudes separately, by calculating the expectation value of only the bosonic or fermionic action with only bosonic of fermionic sources. The $\zeta$ trick then counts only bosonic or fermionic lines. Since QED Feynman diagrams obey \rf{counting} and the additional constraint 
\begin{equation*}
  n_f+2I_f = 2(n_b+2I_b)
\end{equation*}
we can identify $I_b$ and $I_f$ in terms of the number of external lines and the loop order,
\begin{equation}\begin{split}
  I_b &= \frac{1}{2}n_f + l-1, \\
  I_f &= n_b + \frac{1}{2}n_f + 2(l-1), \\
  V &= n_b + n_f + 2(l-1).
\end{split}\end{equation}
This means that the effect of $\zeta$ is still to multiply each $<n>_l$ by a function of only $n$ and $l$. Expanding in the number of fields and the loop order, evaluating the first two terms in \rf{big-op}, the recurrence relations for photon scattering are
\begin{equation}\label{photons}\begin{split}
  &(n+I_b(n,l))\langle A_{\mu_1}(x_1)\ldots A_{\mu_n}(x_n)\rangle_l = \frac{1}{2}\int\!\ud x\,\, (-i\p^2_x\eta^{\mu\nu})\langle A_{\mu_1}(x_1)\ldots A_{\mu_n}(x_n),A_\mu(x),A_\nu(x)\rangle_{l-1} \\
  &+\frac{1}{2}\text{Sym}\sum_{\substack{p=0 \\ l'=0}}^{\substack{p=n \\ l'=l}} \langle A_{\mu_1}(x_1)\ldots A_{\mu_p}(x_p),A_\mu(x)\rangle_l\frac{-i\p^2_x\eta^{\mu\nu}}{p!(n-p)!}\langle A_{\mu_{p+1}}(x_{p+1})\ldots A_{\mu_n}(x_n),A_\nu(x)\rangle_{l-l'}.
\end{split}\end{equation}
The generating functional for purely fermionic amplitudes is 
\begin{equation*}
  Z[\eta,\bar\eta] = \int\pathD(\psi,\bar\psi)\,\,\exp\bigg(\frac{i}{\hbar}S[\bar\psi,\psi]+\frac{i}{\hbar}\int\!\ud x\,\,\bar\eta\psi + \bar\psi\eta\bigg).
\end{equation*}
With this definition it may be checked that
\begin{equation*}\begin{split}
  \hbar^{2n}&\frac{\delta}{\delta\eta^{a_1}(x_1)}\ldots \frac{\delta}{\delta\eta^{a_n}(x_n)}\frac{\delta}{\delta\bar\eta^{b_1}(y_1)}\ldots \frac{\delta}{\delta\bar\eta^{b_n}(y_n)} Z[\eta,\bar\eta]\\
  &=\int\pathD(\psi,\bar\psi)\,\,\bar\psi_{a_1}(x_1)\ldots \bar\psi_{a_n}(x_n) \psi_{b_1}(y_1)\ldots \psi_{b_n}(y_n)e^{\frac{i}{\hbar}S[\bar\psi,\psi]+\frac{i}{\hbar}\int\!\ud x\,\,\bar\eta\psi + \bar\psi\eta}
\end{split}\end{equation*}
so that the logarithm of $Z$ may be expanded as
\begin{equation*}
  W[\eta,\bar\eta] = \frac{1}{\hbar}\sum\limits_{n=0}^\infty \frac{1}{(n!)^2}\int \bar\eta^{b_n}\ldots \bar\eta^{b_1}\eta^{a_n}\ldots \eta^{a_1}\sum\limits_{l=0}^\infty\langle \bar\psi_{a_1}\ldots \psi_{b_n}\rangle_l\hbar^l
\end{equation*}
where the fermionic connected correlation functions have been abbreviated to
\begin{equation*}
\langle \bar\psi_{a_1}\ldots \psi_{b_n}\rangle_l:= \bra{0}\,T\big[ \bar\psi_{a_1}(x_1)\ldots \bar\psi_{a_n}(x_n)\psi_{b_1}(y_1)\ldots \psi_{b_n}(y_n)\big]\ket{0}_l,
\end{equation*}
and the time ordering observes the fermi statistics of the fields. The $\zeta$ trick here multiplies the $2n$-field $l$-loop correlation function by $-2n-I_f(2n,l)$. The only changes to the bosonic case when evaluating the functional derivatives is to keep careful track of the minus signs. We find
\begin{equation}\begin{split}
&(2n+ I_f(2n,l))\langle \bar\psi_{a_1}\ldots\psi_{b_n}\rangle_l = -\int\!\ud x\,\,(-\gamma.\p_x-im)^{ab}\langle \bar\psi_{a_1}\ldots\bar\psi_{a_n}\bar\psi_a\psi_b\psi_{b_1}\ldots \psi_{b_n}\rangle_{l-1} \\
&- \text{ASym}\sum_{\substack{p=1 \\ l'=0}}^{\substack{p=n \\ l'=l}}\int\!\ud x\,\,\langle\bar\psi_{a_1}\ldots \bar\psi_{a_{p-1}}\bar\psi_a\psi_{b_1}\ldots\psi_{b_p}\rangle_{l'}\frac{(-\gamma.\p_x-im)^{ab}}{K_{np}}\langle\bar\psi_{a_p}\ldots \bar\psi_{a_n}\psi_b\psi_{b_{p+1}}\ldots\psi_{b_n}\rangle_{l-l'}
\end{split}\end{equation}
where $K_{np} = (-)^{(n-1)(p-1)}p!(p-1)!(n-p+1!)(n-p)!$ and `ASym' is the instruction to anti symmetrise over the possible external field variables, with the ordering on the LHS given sign $+1$.

Giving the full set of equations for all QED correlation functions is a simple extension, though we will have to reduce our notation further to present the result. Including all the sources in the partition function, we expand $W$ as
\begin{equation*}
W = \frac{1}{\hbar}\sum\limits_{n,m=0}^\infty \int\!R(n,m)J^{\mu_1}\ldots J^{\mu_n}{\bar\eta}^{b_n}\ldots {\bar\eta}^{b_1}{\eta}^{a_m}\ldots {\eta}^{a_1}\sum\limits_{l=0}\langle\,1\ldots n| 1\ldots m| 1\ldots m \rangle_l \hbar^l
\end{equation*}
where $R(n,m) = i^n/\big(n!(m!)^2\big)$ and the correlation functions are 
\begin{equation}
  \langle\,1\ldots n| 1\ldots m| 1\ldots m \rangle_l := \bra{0}T\big[A_{\mu_1}\ldots A_{\mu_n}\bar\psi_{a_1}\ldots\bar\psi_{a_m}\psi_{b_1}\ldots\psi_{b_m}\big]\ket{0}_l.
\end{equation}
Equating by order in $\hbar$ and the number of fields gives
\begin{align*}\label{QED result}
(E + I)\langle\,1\ldots n| 1\ldots m| 1\ldots m \rangle_l = \frac{1}{2}\int\!\ud x\,\,(-i\eta^{\mu\nu}\p^2_x) \langle\,1\ldots n,\{x,\mu\},\{x,\nu\}| 1\ldots m| 1\ldots m \,\rangle_{l-1}\\
-\int\!\ud x\,\,(-\gamma.\p_x-im)^{ab}\langle 1\ldots n | 1\ldots m,\{a,x \}| \{b,x\}, 1\ldots m\rangle_{l-1}
\end{align*}
\begin{equation*}\begin{split}
+\text{Perm}\hspace{-20pt}\sum\limits_{n'=0, m'=1, l'=0}^{n,m,l} \int\!\ud x\langle\,1\ldots n',\{x,\mu\}| 1\ldots m' | &1\ldots m'\,\rangle_{l'} \frac{-i\eta^{\mu\nu}\p^2_x}{2Q_{n'm'}} \\
&\langle n'\ldots n,\{x,\nu\}| m'+1\ldots m| m'+1\ldots m\rangle_{l-l'}
\end{split}\end{equation*}
\begin{equation*}\begin{split}
- \text{Perm}\hspace{-20pt}\sum\limits_{n'=0, m'=1, l'=0}^{n, m, l}\int\!\ud x\,\,\langle 1\ldots n'| 1\ldots m'-1, \{&a,x\}| 1\ldots m'\rangle_{l'}\frac{(-\gamma.\p_x-im)^{ab}}{V_{n'm'}} \\
&\langle n'+1\ldots n| m'\ldots m | \{b,x\}{m'+1}\ldots m\rangle_{l-l'}
\end{split}\end{equation*}
where $E+I =n+2m + I_b+I_f$, $Q_{n'm'} = (-)^{mm'-m'}n'!(n-n')!(m'!)^2(m-m')!^2$, $V_{n'm'}=K_{m m'}n'!(n-n')!$ and `Perm' is the instruction to symmetrise over the bosonic variables and anti symmetrise over the fermionic variables.

\section{Conclusions}\label{Conclusions}
In this paper we have used the cutting and sewing of Feynman diagrams to describe the origins of recurrence relations in the correlation functions and scattering amplitudes they underlie. We have derived recurrence relations for the correlation functions in scalar field theory and quantum electrodynamics which hold to all loop orders. Inserting a resolution of unity into the relations, as in the scalar field case \rf{reso}, generates recurrence relations between scattering amplitudes.

The Schwinger-Dyson equations can be used to give an alternative derivation of our results as we will now describe. The QED Schwinger-Dyson equations for correlation functions of the photon field are
\begin{equation*}
\frac{i}{\hbar}\int\pathD(A,\bar\psi,\psi)\,\,\bigg(\frac{\delta S_0[A]}{\delta A_\mu} + \frac{\delta S_\text{Int}[\bar\psi,\psi]}{\delta A_\mu} + J^\mu\bigg) e^{iS/\hbar + i\int\! J^\mu A_\mu/\hbar}=0.
\end{equation*}
Taking a derivative with respect to the source gives
\begin{equation}\label{SD-1}
\text{Tr}(\eta)\delta(0)Z[J] + \frac{i}{\hbar}\int\pathD(A,\bar\psi,\psi)\,\,\bigg(2S_0[A] + S_\text{Int}[A,\bar\psi,\psi] - i\hbar J^\mu\frac{\delta}{\delta J^\mu}\bigg)e^{iS/\hbar + i\int\! J^\mu A_\mu/\hbar} =0.
\end{equation}
The first term is divergent and should be regulated, but we will see in a moment that it cancels another divergence coming from the free piece of $Z[J]$. Using the $\zeta$ trick from earlier we know that the third term in \rf{SD-1} counts the number of vertices in a correlation function. By expanding $W$ in powers of $J$ it is easily checked that the fourth term in \rf{SD-1} counts the number of external lines. We will evaluate the second term in terms of functional derivatives as we did before, but this time we will isolate the free part of the generating functional, defining $\widetilde{W}$,
\begin{equation}
  W[J] = -\frac{1}{2\hbar}\int J^\mu G_{\mu\nu}J^\nu + \widetilde{W}[J].
\end{equation}
Evaluating the free action in terms of functional derivatives we find that the divergence cancels against a similar term from the free field partition function and the remaining terms depend only on $\widetilde{W}$. The result is
\begin{equation}\label{alt-qed-result}\begin{split}
\hbar\int\!\ud x\,\, &(-i\partial^2_x\eta^{\mu\nu})\frac{\delta^2 \widetilde{W}}{\delta J^\mu(x)\delta J^\nu(x)} + \frac{\delta \widetilde{W}}{\delta J^\mu(x)}(-i\partial^2_x\eta^{\mu\nu} )\frac{\delta \widetilde{W}}{\delta J^\nu(x)} \\
&=  \int\ud x\,\,J^\mu(x)\frac{\delta \widetilde{W}}{\delta J^\mu(x)} - \frac{1}{\hbar} \sum\limits_{n=0}^\infty \int\frac{i^n}{n!}J^n\sum\limits_{l=0}^\infty \hbar^l NV(n,l)\langle\,1\ldots n\,\rangle_l
\end{split}\end{equation}
which is equivalent to our earlier result for the scattering of photons, equation \rf{photons}. The same reasoning with the scalar field directly leads to a tidier version of \rf{final} in terms of $\widetilde{W}$.

Our method also gives us an elegant interpretation of the Schwinger-Dyson equations in perturbation theory. Consider the scalar field Schwinger-Dyson equation,
\begin{equation*}
  \frac{i}{\hbar}\int\pathD\phi\bigg(\frac{\delta S_0[\phi]}{\delta\phi(x)} + \frac{\delta S_\text{Int}[\phi]}{\delta\phi(x)} + J(x)\bigg)e^{iS/\hbar+ i\int\!J\phi/\hbar}=0.
\end{equation*}
Taking a derivative with respect to $J(x)$ gives
\begin{equation}\label{SD-2}
  \delta(0)Z[J] + \frac{i}{\hbar}\int\pathD\phi\bigg(2S_0[\phi]+ N S_\text{Int}[\phi]-i\hbar\int \! J(x)\frac{\delta}{\delta J(x)}\bigg)e^{iS/\hbar+ i\int\!J\phi/\hbar}=0.
\end{equation}
As before the divergent first term cancels a divergence in the free part of $Z$ and we will neglect it. The first term in large brackets counts minus twice the number of internal and external lines, $-2(n+I)$, in a diagram. The second term counts $N$ times the number of vertices, $NV$. The third term counts the number of external lines. Therefore the non-perturbative Schwinger-Dyson equation is equivalent to 
\begin{equation*}
  -2(n+I) + NV + n =0,
\end{equation*}
in perturbation theory, which is just the Euler equation.

The relations we found are of the same, very basic, structure found in the MHV rules of Yang-Mills theory - diagrams for $n$ particle scattering may be separated into diagrams for lower order scattering by severing an internal (non loop) line, or built from $n+2$ particle diagrams of one lower loop order by attaching two external legs together to form a loop.  Although our approach does not give a reason why amplitudes with particular helicities should have special properties, it does suggest that the recurrence relations hold in a much wider context. A issue with our method, however, is that in theories with interactions of multiple orders, such as Yang-Mills and light cone string field theory, the number of vertices and internal lines is no longer unique in the diagrams of a correlation function so the effect of the parameter $\zeta$ no longer has the simple  interpretation of \rf{factor}. In principal the extension is very simple; the expectation value of the whole action gives a counting which is unique in the above sense but requires evaluating the interaction terms as functional derivatives which would give much more complex results. After posting this paper we became aware of \cite{Nair}, where similar arguments with the Schwinger Dyson equations were used to investigate precisely this issue in Yang-Mills theory. The recursion relations between scattering amplitudes given in that paper are similar to our own but contain terms cubic as well as quadratic in the amplitudes which originate in the inclusion of interaction terms. However, simple diagrammatic examples such as that used in chapter 2 can be constructed for theories with interactions of multiple orders. Due to this and the MHV results in the literature we suspect that the effects of the interactions on the recursion relations may be largely degenerate. We will address this in a future publication; in particular we will consider the scattering of strings as described by string field theory.

\subsection*{Acknowledgements}
Many thanks to Paul Mansfield for useful discussions and to James Gray, Emily Hackett-Jones, Vishnu Jejjala and David Leonard for proof reading and helpful comments.

\end{document}